 %
% >>> SPELLING
%

\input psfig.sty
\magnification=\magstep1
\hsize=6truein
\overfullrule=0pt

\font\title=cmr10 scaled\magstep5
\font\subtitle=cmr10 scaled\magstep2
\font\author =cmr10 scaled\magstep1
\font\eaddr=cmtt10
\font\ninerm=cmr8
\baselineskip=16pt plus 1pt minus 1pt
\def\QED{\vrule height8pt width4pt depth0pt}

\def\proclaim #1: #2\par{\medbreak
	      \noindent{\bf#1:\enspace}{\sl#2}\par\medbreak}
\pageno=1	      

\hfill 19/9/2009\break

\vskip 1truein
\centerline{\title Archimedean Ice}
\vskip .7truein
\centerline{{\author Kari Eloranta}\footnote{$^*$}
{Research supported by the Finnish Academy of Science}}
\vskip .3truein
\centerline{\author Institute of Mathematics}
\centerline{\author Helsinki University of Technology}
\centerline{\author FIN-02015 HUT, Finland}
\vskip .1truein
\centerline{\eaddr kve@math.hut.fi}

\vskip .3truein

\vfill
\eject
\centerline{\subtitle Abstract}
\vskip .4truein
\centerline{\vbox{\hsize 4.5in \noindent \ninerm \strut The striking boundary dependency (the Arctic Circle phenomenon) exhibited in the ice model on the square lattice extends to other planar set-ups. We present these findings for the triangular and the Kagom\'e lattices. Critical connectivity results guarantee that ice configurations can be generated using the simplest and most efficient local actions. Height functions are utilized throughout the analysis. At the end there is a surprise in store: on the remaining Archimedean lattice for which the ice model can be defined, the 3.4.6.4. lattice, the long range behavior is completely different from the other cases.}}

\vskip .6truein
\centerline{\vbox{\hsize 4.5in \noindent Keywords: Ice model, vertex model, Archimedean lattice, probabilistic cellular automaton. 
\vskip .2truein
\noindent AMS Classification: 05B45, 52C20, 68Q80, 82B20, 82C20.
\vskip .2truein
\noindent Running head: Archimedean Ice}}

\vfill
\eject

\hfill{ }\break
\vskip .2truein
\noindent {\subtitle Introduction}
\vskip .2truein

\noindent Although the best known version of the ice model is that defined on the square lattice ([B], [L]), the construction is quite natural on a number of other lattices as well. In this note we investigate the basic properties of the bounded version of the model on the other Archimedean lattices: the triangular, Kagom\'e and 3.4.6.4. lattices. The investigation serves two purposes. Firstly it complements the various studies of the infinite unbounded models and answers the question on the influence of the boundary posed already by Kasteleyn ([K]). Secondly we hope to contribute whatever is possible to the unification of the theories of lattice statistical mechanics, higher dimensional symbolic dynamics and tilings that has been worked on for some time now (starting from [CL], [T], for later development see e.g. [KZ-J]).

Our results show that 18/20/36-vertex models (ice on Kagom\'e, triangular and 3.4.6.4. lattices respectively) can to a certain extent be analyzed with similar means than the square lattice one (the six-vertex model). They have analogous cycle structure which facilitates the configuration computation with simple and efficient algorithms. Height works in the same way in these models and the boundary effects it forces are qualitatively similar -- up to a point. There is a sharp a demarcation of temperate and frozen subdomains akin to the Arctic Circle Phenomenon in dominoes ([JPS]) in the triangular and Kagom\'e set-up. But just as in the context of e.g. the hard square/hexagon model ([B]) there is a surprising lattice dependency already within the set of the four possible Archimedean lattices. Indeed the influence of the underlying lattice is even more pronounced here: ice on 3.4.6.4. lattice is in terms of long range order a qualitatively different model. Unlike the other ice models this one shows strong uniformity in the configurations independent of the boundary arrangement of the arrows.

Whenever analyzing the bounded 6/18/20/36-vertex models, similarities admitting, we refer to the results already established for the six-vertex model ([E]) and concentrate here on the novel features. We make however an effort to make this paper self-contained so that the reader can grasp the main ideas and results already from here.

\vfill\eject

\vskip .3truein
\noindent {\subtitle 1. Set-up}
\vskip .2truein

\noindent In the {\bf vertex-models} of the Statistical Mechanics instead of spin variables one deals with arrow orientations between nearest neighbor lattice sites. The global ensemble of the orientations defines the configuration. 
\proclaim Definition 1.1.: A {\bf vertex configuration} is the arrangement of arrows arriving and departing from a lattice point. It is {\bf legal} for the {\bf ice rule} if there is the same number of incoming and outgoing arrows at that lattice point. If there is a legal vertex configuration at every vertex of the lattice the arrow configuration is legal for the {\bf ice model}. \par

\noindent In the square lattice case there are six such vertex configurations, hence the ice model on that lattice is also called the {\bf six-vertex rule}. The term \lq\lq ice\rq\rq\ stems from the physical interpretation for this model (see [L]). Although this physical interpretation does not carry over to other lattices, for simplicity we call those rules {\bf ice-type}. For the purposes of this paper we only consider ice on planar lattices.

A rule of this kind obviously requires even vertex degree. Among the three regular planar lattices -- the square, triangular and hexagonal lattices -- ice model can be defined on the first two. The next simplest planar lattices are the {\bf Archimedean} or {\bf uniform lattices}. They are defined via tilings: their bonds correspond to the tile edges of such tilings by regular polygons which are up to rotation identical at each vertex. There are 21 ways of tiling a vertex neighborhood with regular polygons. 11 of these arrangements extend to the plane - these are the Archimedean tilings (for the complete list see [GS]). Among these we have four lattices with even vertex degree: square, triangular, Kagom\'e and 3.4.6.4. lattices (the code number $n_1.n_2.n_3.n_4.$ lists the $n$-gons that one sees turning once around a vertex). In physical terms these are the simplest discrete planar structures that accommodate the dipole/incompressibility restriction of the ice-type.

The available vertex configurations are illustrated in Figure 1. Triangular lattice is on the left with multiplicities accounting rotations and reflections listed below. In the center we have the analogous arrangements for the square lattice. The multiplicities for Kagom\'e and 3.4.6.4. are obtained from those of the square lattice vertex configurations by multiplying them with the numbers on the right, below (the numbers of possible orientations of the middle arrangements in these lattices). Adding the numbers up we could call these 20/6/18/36-vertex models (for triangular, square, Kagom\'e and 3.4.6.4. lattices respectively) but for simplicity we just call them ice models on the appropriate lattice.

\vskip .3truein

\centerline{\hbox{\psfig{figure=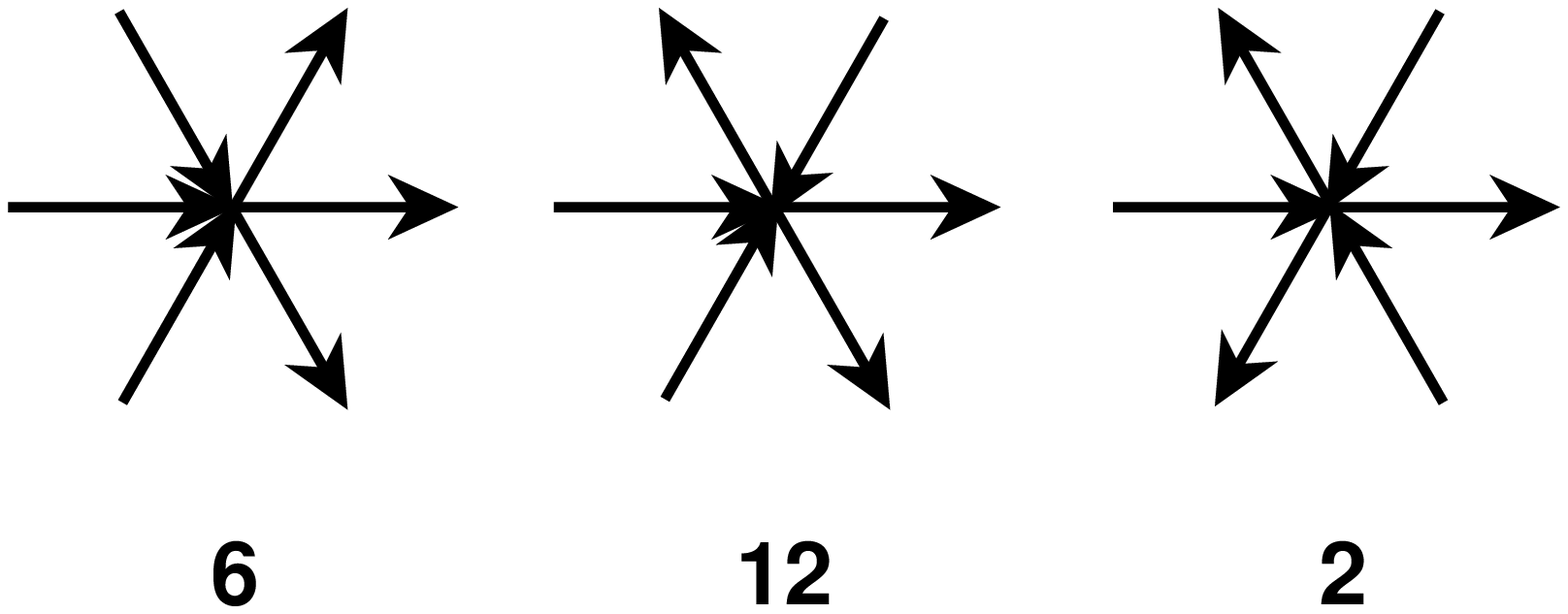,height=.8in} 
\hskip .3in\psfig{figure=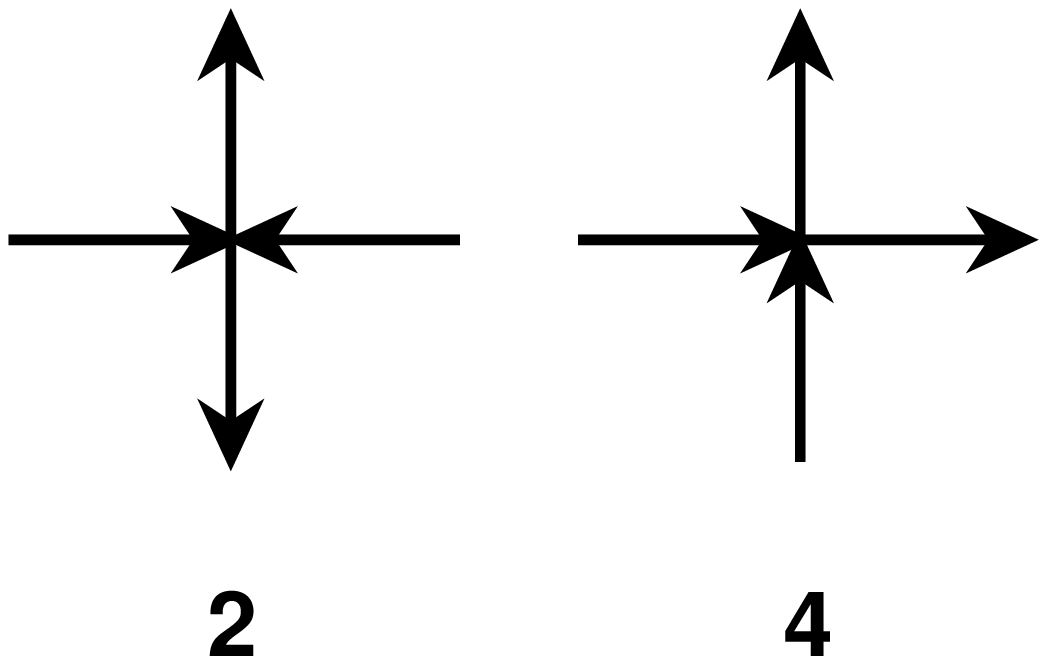,height=.8in}
\hskip .3in\psfig{figure=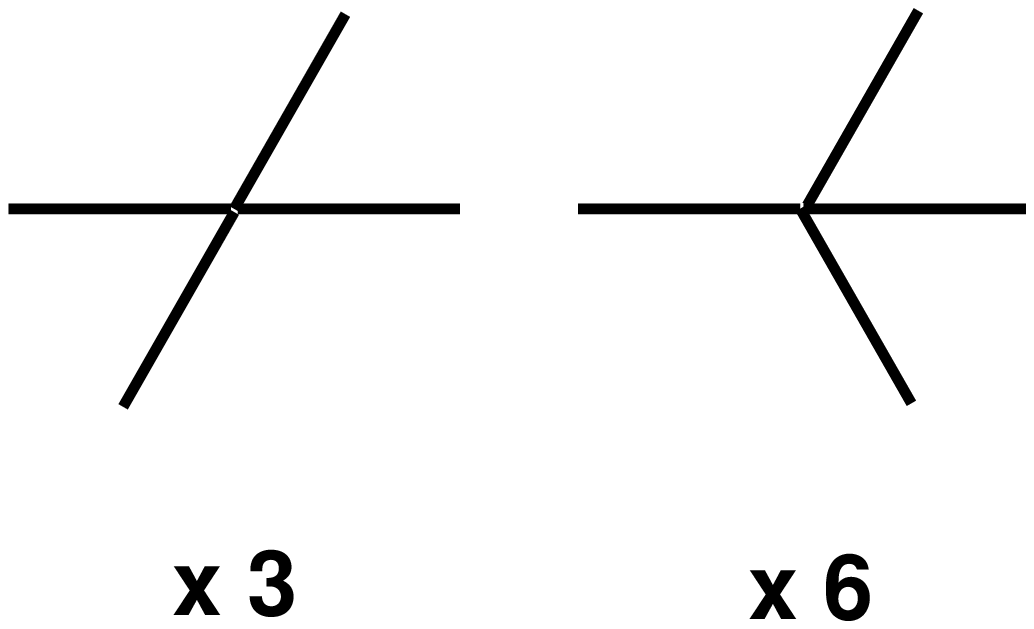,height=.8in} } }
\vskip .1truein
\noindent {\bf Figure 1a, b, c.} Vertex configurations: arrow arrangements and multiplicities.

\vskip .2truein
\noindent The square ice on a bounded domain was considered in an earlier paper ([E]). Here we study the other three models on a hexagonal domain. A $N${\bf -hexagon} in the case of the triangular lattice is a domain which is oriented along the lattice axes with $N$ boundary arrows along each edge, $N$ even. Or equivalently we can require that along each edge there are $N/2$ lattice sites with all six arrows attached to them. Figure 2a. illustrates the area around leftmost corner of such hexagon (the boundary will have six-fold symmetry). The {\bf boundary arrows}, $6N-6$ in total, are rendered bold. They will be fixed and the main problem will be determining when and how the interior arrows (lighter) can be arranged into a legal configuration.

The other simple domain shapes on the lattice, a unilateral triangle and a rhombus turn out to be somewhat restrictive due to the acute corners. On the hexagonal domain we will be able to illustrate both \lq\lq ordered/frozen \rq\rq\ and \lq\lq disordered/temperate\rq\rq\ configurations and their coexistence as in a diamond in the square lattice case in the preceding study ([E]).

The dual lattice of the triangular lattice is the hexagonal lattice. Every lattice site is the center of a minimal (unit) hexagon the boundary of which we should think of having a clockwise orientation. By the ice rule the total {\bf flux} across this boundary is zero (ingoing arrow counts $+1$, outgoing $-1$). Consider the maximal dual lattice loop on the domain, {\bf the boundary loop} (the dual lattice edges of this loop still cross arrows in the $N$-hexagon). It is the sum of all the directed unit hexagons in the dual lattice inside the domain. Hence if the configuration inside the domain is legal then the flux across the boundary loop vanishes.

\vskip .2truein
\noindent In the Kagom\'e-lattice $N$-hexagon has $N$ lattice sites and $N/2$-arrows along each edge. Figure 2c. illustrates the leftmost corner of such hexagon. The bold arrows are again the fixed boundary arrows. Because of the ice-rule the flux around each lattice point vanishes. Therefore a legal fill-in of a hexagon will have zero boundary flux. Same principles extend to the 3.4.6.4. lattice case.

\vskip .3truein
\centerline{
\vbox{\hbox{\psfig{figure=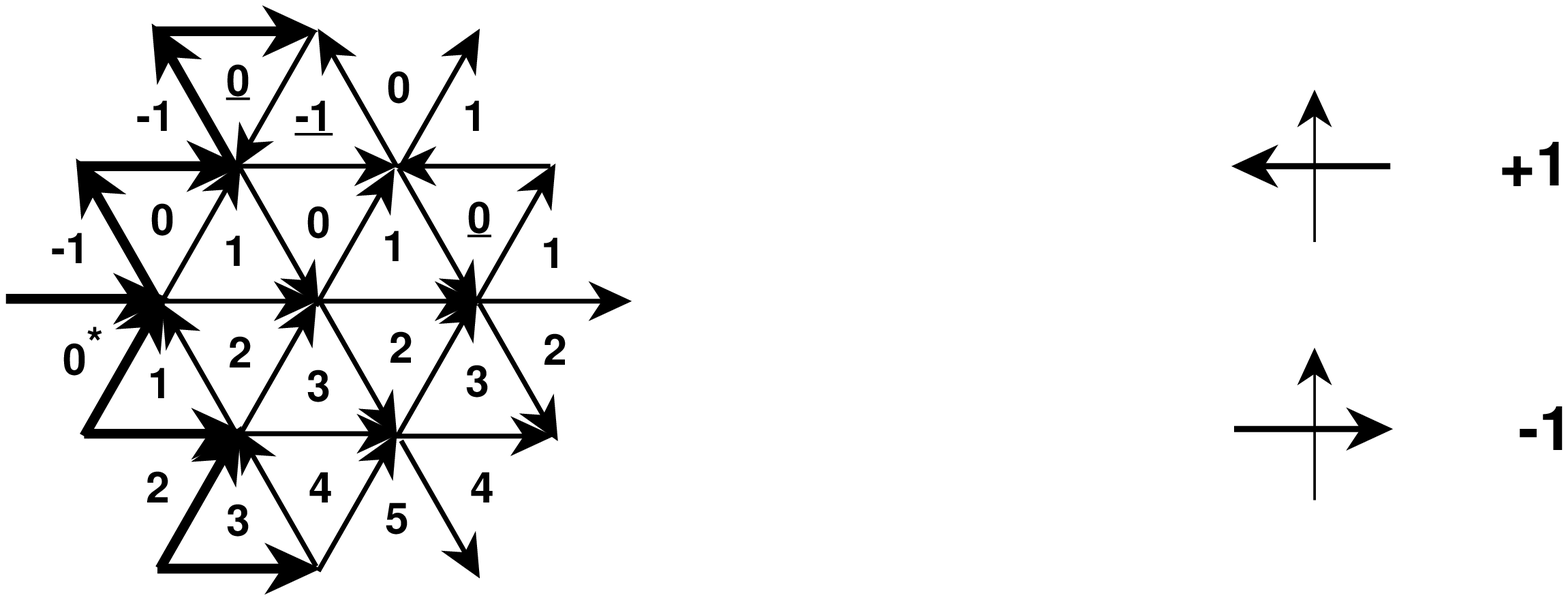,height=1.2in} }
\vskip .1in {\hbox{\psfig{figure=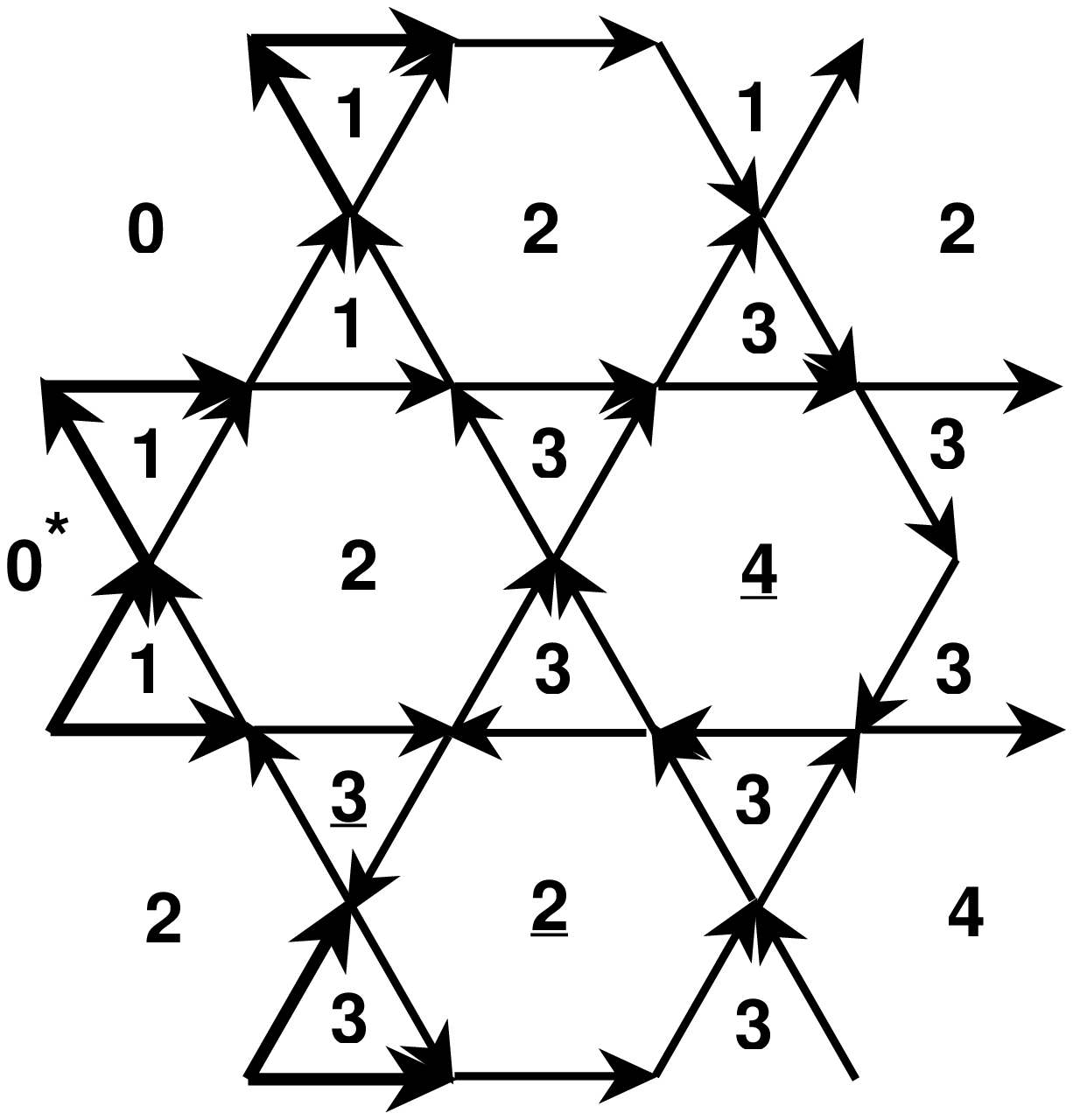,height=1.3in}
\hskip .5in \psfig{figure=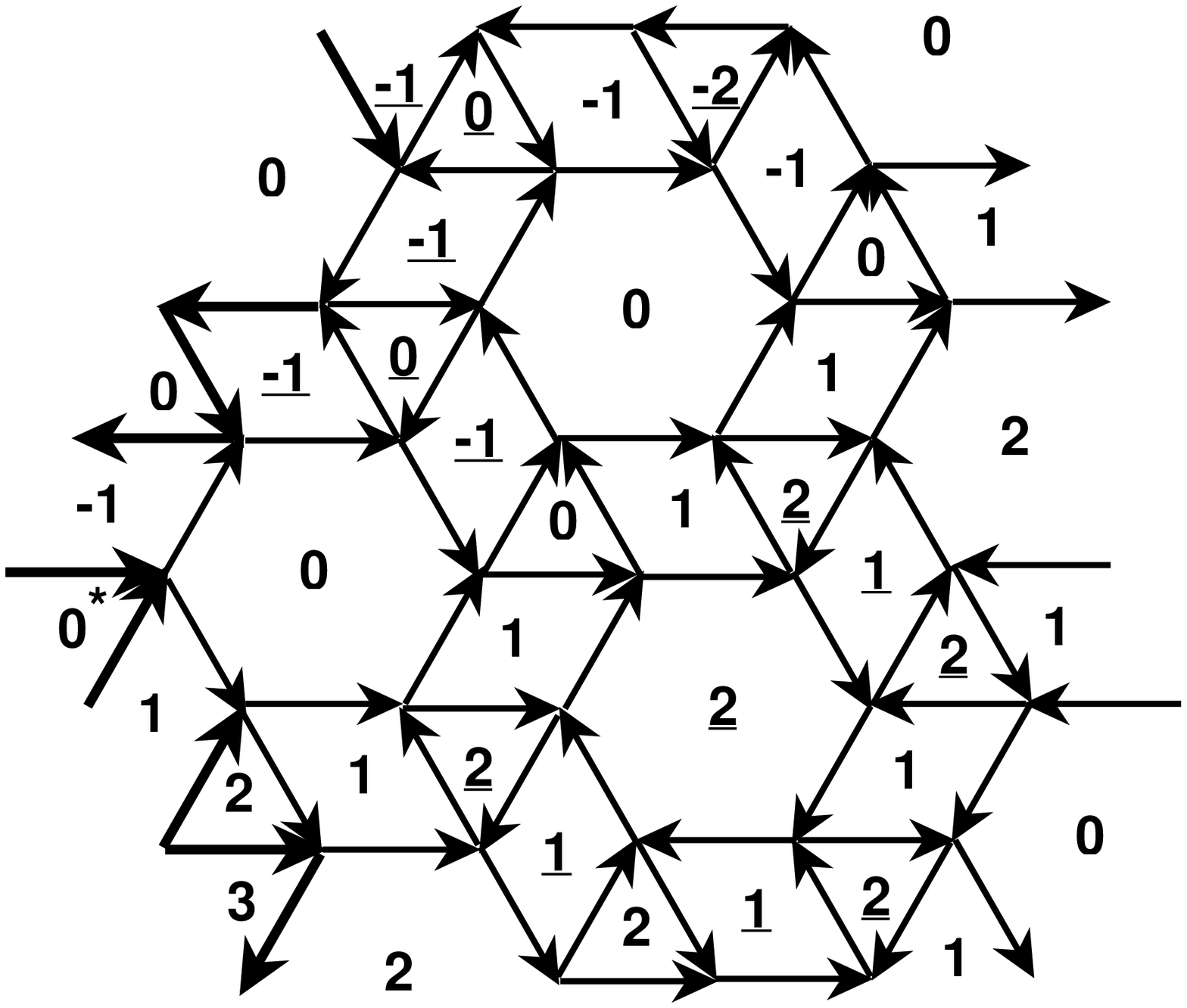,height=1.8in} } } }
}
\vskip .2truein
\noindent {\bf Figure 2a, b (top), c, d (bottom).} The crossing rule and configurations with heights.
\vskip .2truein

\noindent Let $C$ be the set of legal ice configurations on a given triangular lattice hexagon and let $D$, the {\bf dual cover}, denote the finite subset of the hexagonal lattice that has the property that an edge connecting a nearest neighbor pair of vertices from $D$ crosses an arrow $c\in C.$

The {\bf height}, $f:C\times D\rightarrow {\rm {\bf Z}}$, is an extremely useful function in analyzing ice-type models. Its increments on $D$ are defined by the crossing rule in Figure 2b. The light arrow indicates the edge on the dual lattice that we move along and the bold marks the configuration arrow. The rules apply in all possible rotations. Note that height around a closed loop in $D$ vanishes (since this is the same as computing the flux across that loop). Hence $f$ is independent of the path along which it was computed. To be unique it needs to be specified at one {\bf base point} which we choose to be the leftmost dual lattice point (starred). In Figure 2a. we have indicated the heights with the choice that at the base point height vanishes. 

The dual of the Kagom\'e/3.4.6.4. lattice is the rhombus-lattice [3.6.3.6]/[3.4.6.4] respectively. The definition of height on them is as above and we have indicated the values in Figure 2c. and 2d (The numbers in square brackets refer to Laves tilings, see [GS]. Note that 3.4.6.4. being here rendered so that squares turn into lozenges does not in any way affect arguments since the neighborhood topology remains intact).

\vskip .2truein
\noindent The three cases when height increases/decreases at a maximal rate or alternates in value along a path in the dual lattice will be important in later considerations. As the boundary specification will be critical we have also chosen to illustrate this on the boundaries in Figure 2. The boundary arrows in the lower halves of the samples are of maximal {\bf tilt} (discrete derivative) as we trace the boundary. The upper half of the boundary arrows illustrate the alternating case. Subsequently we refer these special arrangements as having tilt $\pm 1$ or $0.$

\vskip .3truein
\noindent {\subtitle 2. Connectivity}
\vskip .2truein

\noindent The ice-type local arrow parity expressed in Definition 1.1. has far reaching consequences on the global structure of the configuration ensembles. We now elaborate on this utilizing as well as extending the results in the square lattice case.

\vskip .2truein
\noindent Consider a legal triangular, Kagom\'e or 3.4.6.4. ice configuration. Suppose that we can find a closed unidirectional path of configuration arrows in it (or a path from infinity to infinity). Reversing this directed cycle i.e. flipping every arrow on the cycle results in an other legal configuration since the rule at each vertex is respected. Existence of an unidirectional cycle is therefore related to the non-uniqueness of the fill-in: a boundary arrangement of arrows that allows a fill-in which has an off-boundary unidirectional cycle allows in fact multiple fill-ins.

\proclaim Definition 2.1.: Call the smallest lattice triangle a {\bf 1-triangle}. The orientation $\bigtriangleup$ is {\bf even} and $\bigtriangledown$ is {\bf odd}. On the Kagom\'e lattice we also have a minimal hexagon, a {\bf 1-hexagon} and on the 3.4.6.4. lattice we additionally have left-leaning, right-leaning and straight standing {\bf 1-lozenges}. If these directed {\bf 1-polygons} are unidirectional we call them {\bf 1-cycles} and their reversals {\bf local moves}.\par 

\noindent In the case of a finite cycle one can show by utilizing flux as in the square lattice case (see [E]) that in fact 

\proclaim Proposition 2.2.: A unidirectional cycle always encloses a 1-cycle. \par 

\noindent We say that a cycle is {\bf off-boundary} if it does not contain any of the (fixed) boundary arrows. Define a bounded {\bf frozen} configuration to be one without off-boundary 1-cycles. Its opposite is the {\bf temperate} configuration which we define as one having a directed cycle boundary. 

In Figure 2a, c and d. the sample configurations contain 1-cycles -- they are the ones with the height at the center underlined. The samples are typical in the sense that neither has a cycle boundary yet they have 1-cycles. Except for special boundary configurations the fill-ins will exhibit a coexistence of frozen and temperate subdomains.

Related to the 1-cycles there is a simple but useful notion which we will need in the proof below. In Figure 3a, b. the infinite wedges $C_i$ rooted at the vertices of the 1-cycles are called {\bf contact sectors}. For the triangle oriented upside down we reflect the wedges and for 1-hexagon we have for clarity indicated only the odd sectors.

\vskip .3truein
\centerline{\hbox{
\psfig{figure=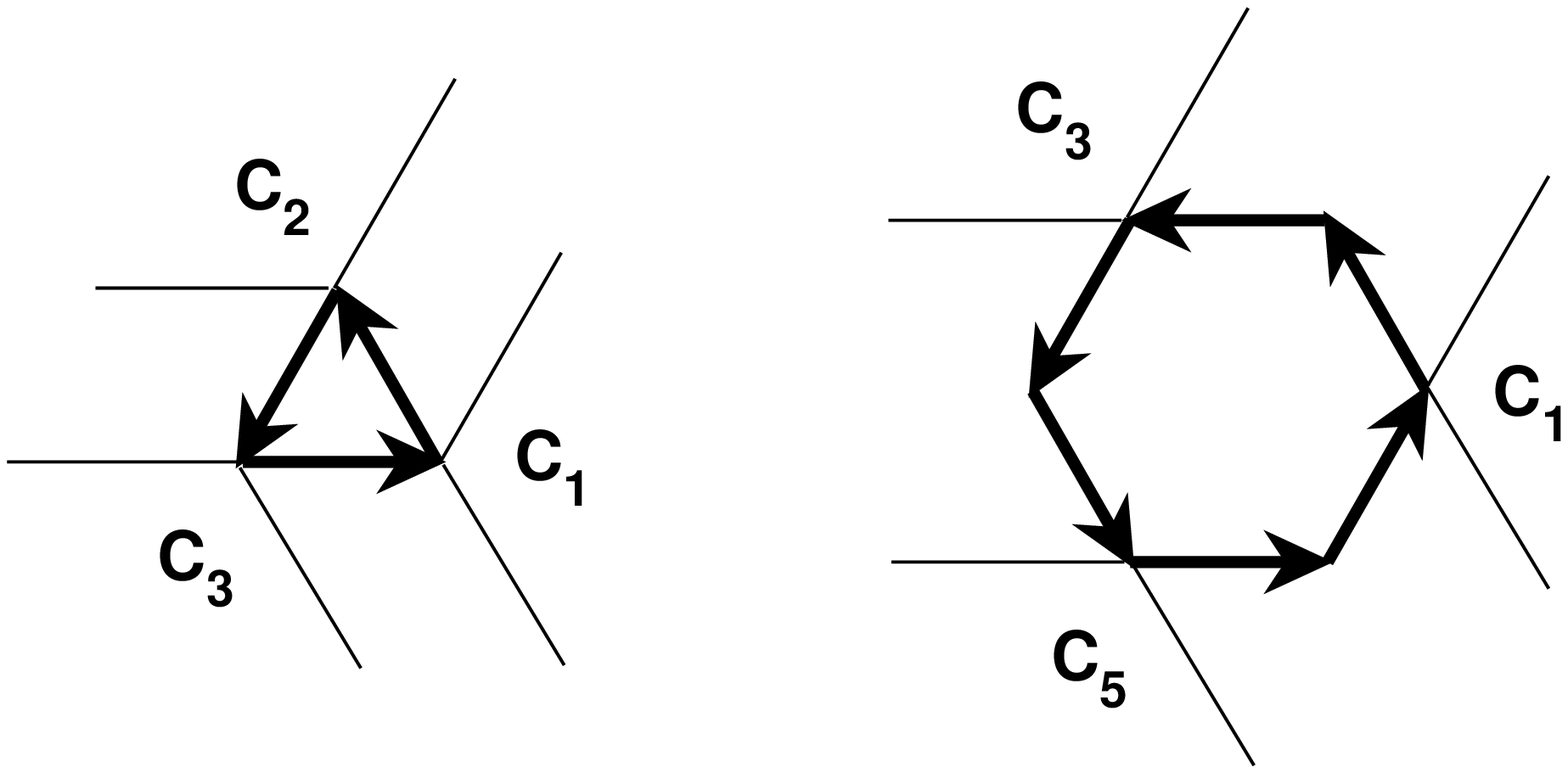,height=.6in}
\hskip .3in
\psfig{figure=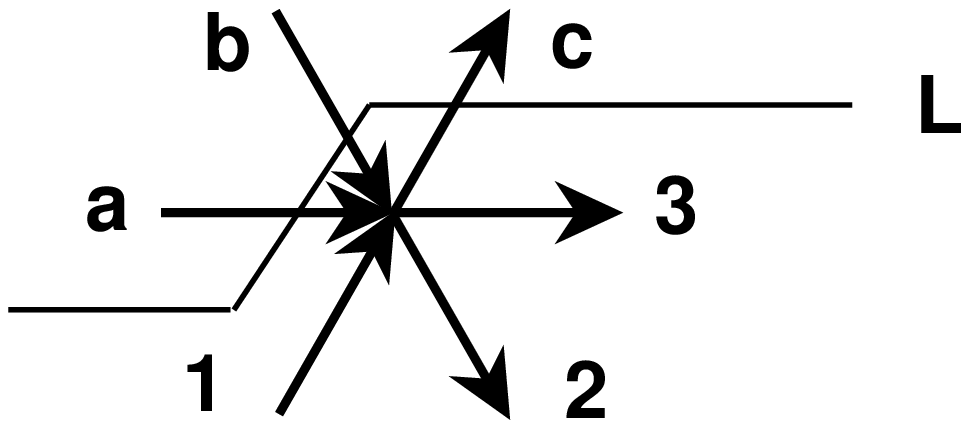,height=.6in} 
\psfig{figure=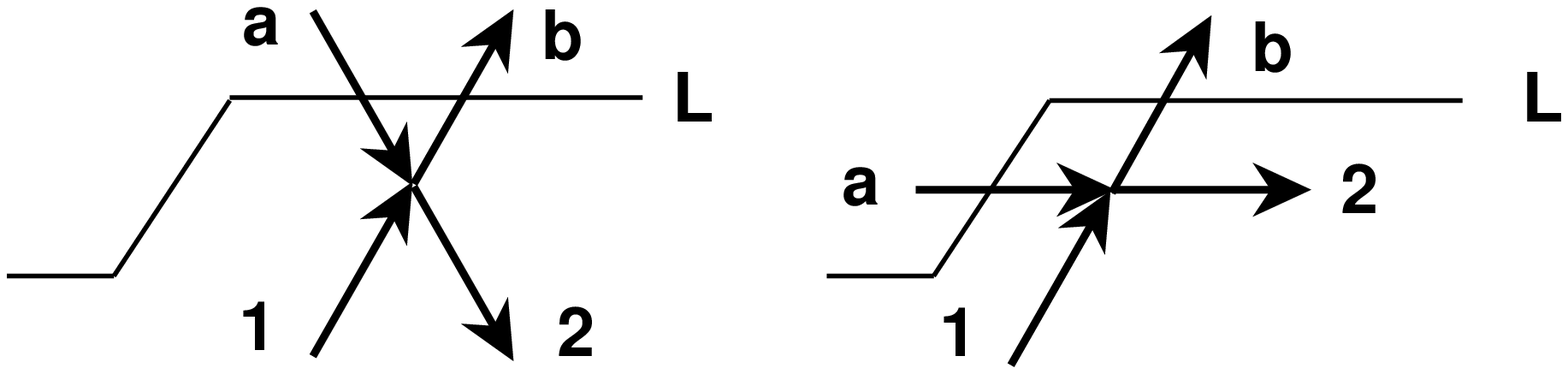,height=.6in} } }

\nobreak
\vskip .1truein
\noindent{{\bf Figure 3a, b.} Contact sectors. {\bf c, d, e.} Vertex configuration mismatch.}
\vskip .2truein

\noindent The cycle structure leads to the connectivity/irreducibility in the following sense.

\proclaim Theorem 2.3.1.: On triangular and Kagom\'e lattices the set of ice configurations on with common boundary arrows on a $N$-hexagon is connected under 1-cycle reversals i.e. two such configurations can be transformed to each other with a finite sequence of 1-cycle reversals. \par

\noindent {\bf Proof:} The argument uses a \lq\lq lexicographic sweep\rq\rq\ as in the square case, now with some refinements. When during the sweep we arrive to a lattice point $l$ where there is the next mismatch between the two configurations under comparison, the situation in triangular lattice looks like in Figure 3c. $L$ denotes the \lq\lq front\rq\rq\ above which all vertex configurations in the two configurations match. In the Kagom\'e case we may encounter three different arrangements, two of which are in Figures 3d, e. (the third is like the rightmost, but rotated 60 degrees clockwise).

Consider first the case on triangular lattice. The three arrows $a-c$ cannot all be in or all out since in that case there cannot be a mismatch. So among arrows $1-3$ there is a 2-1 or 1-2 division between ingoing and outgoing arrows. Hence we can always find among them a pair ($(1,2)$, $(1,3)$ or $(2,3)$), same pair on both configurations, so that the arrows in the pairs are unidirectional but oriented opposite in the two configurations. One then extends these 2-paths to closed off-boundary cycles, same cycles but with opposite orientations, on the two configurations. 

Pick one of the configurations e.g. the one with clockwise oriented cycle, $O_1.$ By Proposition 2.2. inside it there is at least one 1-cycle of some type. Denote their collection by $\left\{C_i\right\}.$ Choose two of its contact sectors in such a way that they do not overlap and do not contain the point of mismatch $l.$ One can easily show that it is possible to find two directed paths, one in each of the contact sectors, which connect the 1-cycle to $O_1.$ Moreover the orientations of these paths are such that a new clockwise directed cycle is formed that passes through $l$, along the edge of the 1-cycle and is contained in the domain bounded by $O_1.$ This construction is done to all of the 1-cycles inside $O_1.$ Finally define the natural minimal directed cycle along these new cycles inside $O_1$ and call it $\tilde O_1.$

Now some the 1-cycles inside $O_1$ are on the boundary of the $\tilde O_1$ and none are strictly in its interior. By reversing these cycles if necessary we obtain a new directed cycle $O_2$ with the property that all of the 1-cycles $\left\{C_i\right\}$ are left outside it. Moreover $O_2$ encloses a strictly smaller area than $O_1$, all inside it. 

Applying the argument above to $O_2,\ O_3$ and so on finally forces a 1-cycle that has $l$ as its vertex. Hence we correct a mismatch at $l.$ Note that if the path $O_1$ involved the pair $(1,3)$ we need a second application of this argument, now to a loop through $(1,2)$ or $(2,3)$, whichever pair is still mismatched in the two configurations. After this all arrows at $l$ match and the front $L$ moves to the next lexicographic location to check for a mismatch.

For the Kagom\'e lattice the argument is similar. Whether the 1-cycle is a 1-triangle or 1-hexagon makes no difference except in the choice of the contact sectors. Note that for Kagom\'e lattice (as well as square lattice) one pass of the argument above for each $l$ is sufficient to correct the mismatch. \hfill\QED

\noindent {\bf Remark:} Since the \lq\lq sweep\rq\rq\ is a rather general procedure to deal with the configuration, the result will hold for more general simply connected domain geometries as well. But since our presentation is geared toward analyzing boundary dependency in a hexagonal domain we refrain from pursuing this.

\vskip .2truein
\noindent The argument above is the most direct way that we know of to argue the connectivity for the set of ice configurations and that is why we present it here. The result however is more general and in particular holds for the 3.4.6.4. lattice as well. Since the contact sectors do not seem to work in this context we now briefly sketch a more general way of arguing the result (the most general, abstract argument can be found in [P]).

For all our four lattices the height function introduces a partial order in the set of configurations with common boundary configuration. We say that $c_1$ majors $c_2$, $c_1\succeq c_2$ if $f(c_1, d)\ge f(c_2, d)$ at every point $d$ on the dual lattice. The local minima of the height surface for a given configuration are simply the center points of counterclockwise oriented 1-cycles. If these are off-boundary one can reverse them and reach a strictly higher height surface. After a finite number of steps one arrives at the maximal element $\overline c$ (such that $f({\overline c}, d)\ge f(c, d)$ for all $d$ and any $c$ with the same boundary configuration). The maximal element has by definition no off-boundary counterclockwise oriented 1-cycles. Through this (or the minimal element) one can connect all the configurations with a finite sequence of local moves. This argument leads to

\proclaim Theorem 2.3.2.: The set of 3.4.6.4. ice configurations with common boundary arrows on a $N$-hexagon is connected under 1-cycle reversals. \par

\vskip .1truein
\noindent The results above are optimal: no smaller set of elementary moves will guarantee connectivity. This will now be shown through counterexamples.

Suppose that we have a configuration on the triangular lattice where each of the lattice arrows is directed either toward 1, 3 or 5 o'clock. Then reverse all the arrows on one of the 1 o'clock and 5 o'clock lattice lines. Cut a $N$-hexagon out from this so that the intersection of these lattice lines is at the center. The patch that we see at the center of the hexagon looks like Figure 4a.

Reversing the 1-triangle at $C$ will obviously not affect the boundary arrows i.e. we get another configuration, call it $\tilde H$, compatible with the boundary configuration. If we are subsequently only allowed to act with $\bigtriangleup$-reversals there will be only two such triangles to work on, the one at $T$ in $\tilde H$ and the one below $C.$ But it is easy to see that reversing these and any other directed $\bigtriangleup$-cycles will never yield a directed cycle outside the two acute wedges defined by the bold lines. Hence two sides of $C$ will never be returned to their original orientation. 

In the case of Kagom\'e lattice we generate a configuration as above, this time from arrow lines pointing toward 3, 7 and 11 o'clock. This configuration has has the property that all 1-triangles are directed (see Figure 4b.). If we reverse any one of them, say $C$, the new configuration is compatible with the boundary but still has no directed 1-hexagons. Notice that even if only the reversal of $\bigtriangledown$ is forbidden we are still stuck. Reversing $C$ cannot be undone with $\bigtriangleup$-actions and all the 1-hexagons will still remain undirected.

If the configuration is generated in an alternating fashion from arrow lines to directions 1 and 7, 3 and 9 and 5 and 11 o'clock we can see in it a patch like Figure 4c. If we reverse the 1-hexagon, the 1-triangles around it will become directed but the David's star is isolated from the 1-cycle reversals outside it, so its original orientations cannot be returned.

For the 3.4.6.4. lattice consider the arrangement on the right, Figure 4d. This hexagon clearly extends to a unique global configuration. Suppose we reverse the even triangle marked with $C$ and then ban subsequent even 1-triangle reversals. It is easy to see that the right-leaning lozenges can never be reversed in this configuration. Hence return to the original configuration is impossible. Similarly if after this move we reverse the left-leaning lozenge at $L$ and then ban left-leaning lozenge moves, the original configuration cannot be recaptured with the remaining moves.

Finally if we pick a configuration with only 1-cycles, flip one of the 1-hexagons and then forbid 1-hexagon moves, no neighboring lozenges can be reversed and the original configuration cannot be recaptured. 

\vskip .3truein
\centerline{\hbox{\psfig{figure=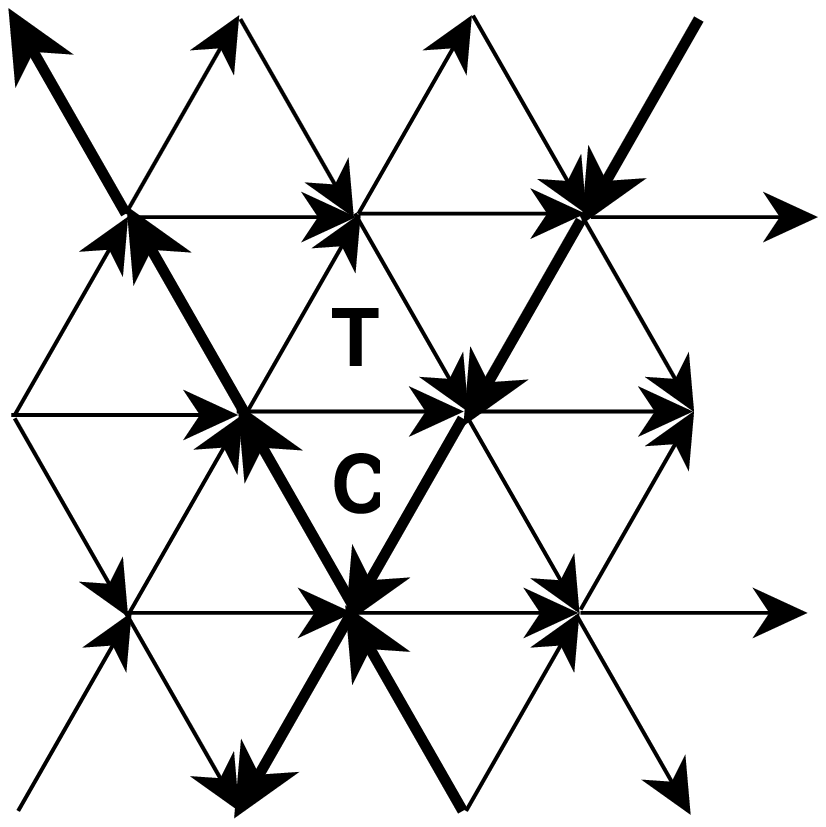,height=1in} 
\hskip .3in\psfig{figure=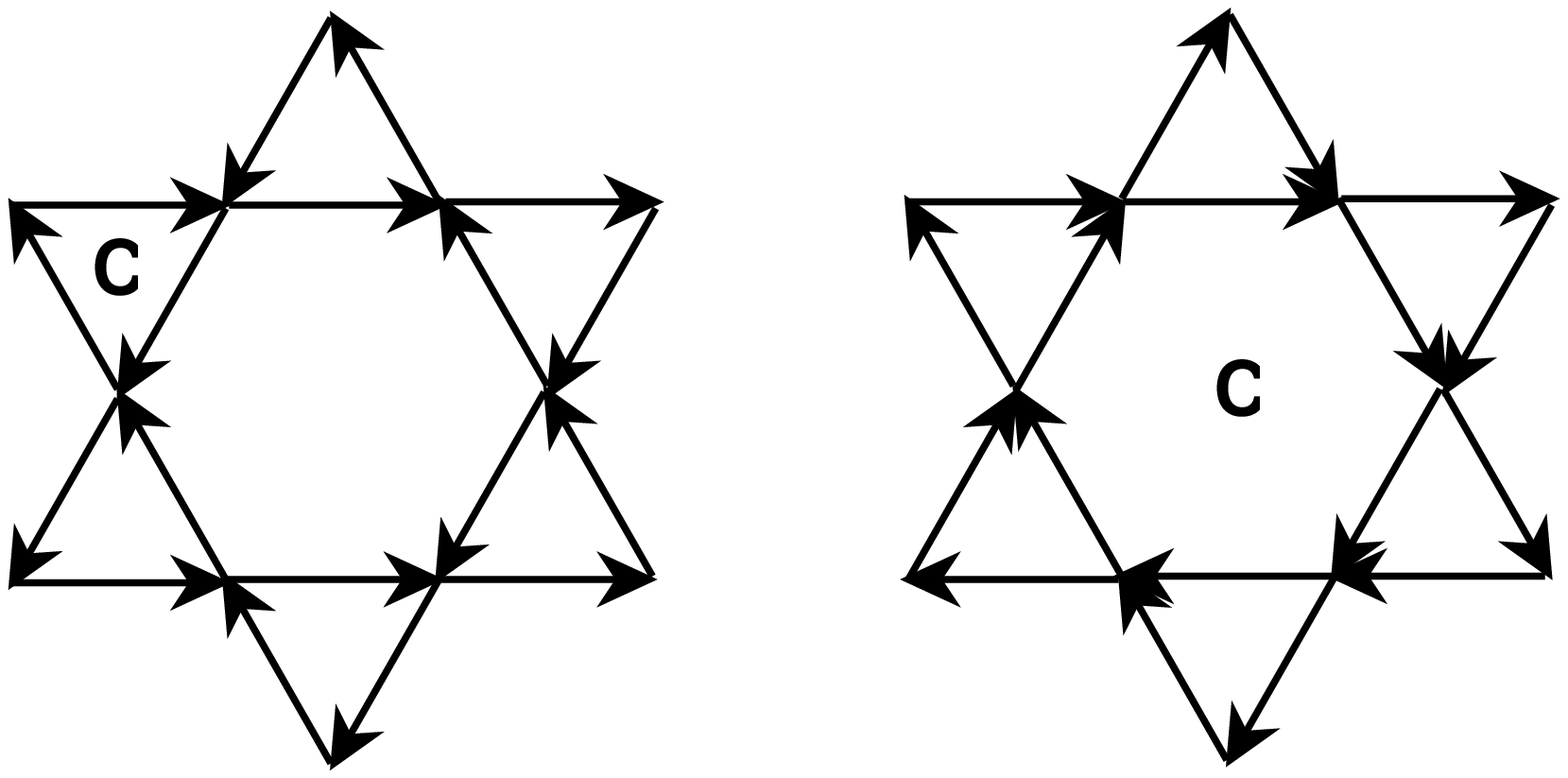,height=1in}
\hskip .3in\psfig{figure=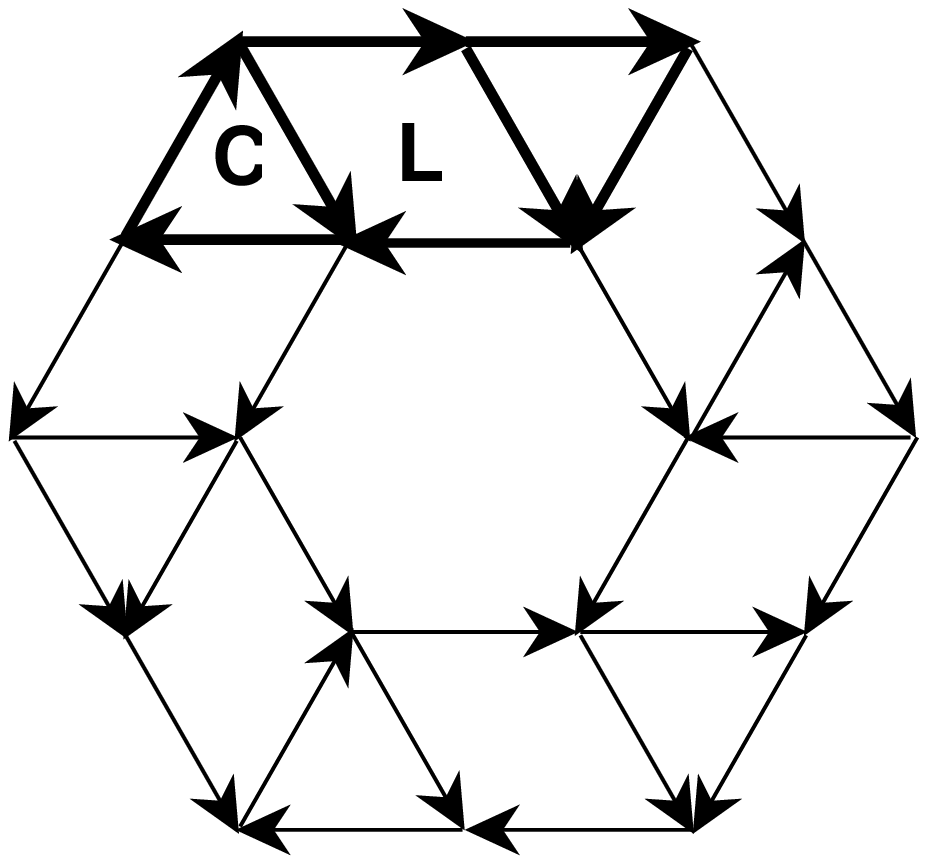,height=1in} } }
\vskip .1truein
\noindent{\bf Figure 4a, b, c, d.} Restricted actions.
\vskip .2truein

\noindent From these counter examples we conclude a slight sharpening to the Theorems 2.3.x. above.

\proclaim Proposition 2.4.: The connectivity results fail if not all 2/3/6 types of local moves are available for triangular/Kagom\'e/3.4.6.4. lattices respectively.\par

\vskip .3truein
\noindent {\subtitle 3. Generating configurations}
\vskip .2truein

\noindent From the practical point of view the most significant consequence of Theorems 2.3.x. is that they facilitate the generation of the configurations with a given boundary configuration. We now briefly indicate how this is done. The aim here is simply to get an algorithm to study the model, not runtime considerations or ultimate efficiency.

\vskip .2truein
\noindent The arrows of the triangular lattice can be viewed as an array of either just even or just odd oriented 1-triangles. Call their restrictions to the $N$-hexagon $L_e$ and $L_o$ respectively.

On the triangle arrays we define two local update rules. If an off-boundary even/odd 1-cycle is encountered, it is reversed independently with probability 1/2 and also its odd/even nearest neighbor 1-triangles are updated accordingly. These rules immediately give global random maps $F_e$ and $F_o$ which check 1-cycles on $L_e$ and $L_o$ respectively and independently update the arrow configuration on both. They define a {\bf probabilistic cellular automaton (pca)} action on the configurations.

On the Kagom\'e lattice we additionally define a local rule reversing an unidirectional off-boundary 1-hexagon with probability 1/2 and updating the six neighboring 1-triangles. Call the global map of independent flips $F_h.$ The updating sequence $\left\{F_e,F_o,F_h\right\}$ defines a pca cycle. In the 3.4.6.4. lattice case we have to perform the lozenge flips as well. Let $F_l$ $F_r$ and $F_s$ denote the probability 1/2 flips of the left leaning, right leaning and straight standing off-boundary lozenges respectively. The updating sequence $\left\{F_e,F_o,F_s,F_l,F_r,F_h\right\}$ has been used in the generation of the configurations of this ice model.

\vskip .2truein
\noindent For a given boundary condition the set of legal fill-ins and the local moves constitutes a finite Markov Chain. Suppose (as above) that all possible transitions have positive probability. Theorems 2.3.x. imply the irreducibility of this chain. Because of the positive transition probabilities the chain is also aperiodic. Hence from the general theory we conclude that the chain is ergodic: given any initial configuration the pcas above will eventually generate any configuration compatible with the boundary configuration of the initial one and weight it uniformly (according to the measure of maximal entropy)([W]). The choice of probability 1/2 for the flip is for maximal speed. Note that in each of the lattices the densities of the 1-$n$-gons of all types are the same (in the set of all 1-polygons). Hence in the update schemes above all local moves are weighted equally. We do not know of any rigorous relaxation rate result applicable here but in all simulations performed it seemed high and is likely to be exponential.

\vskip .3truein
\noindent {\subtitle 4. Boundary dependency}
\vskip .2truein

\noindent We now investigate the key feature of the finite versions of lattice models -- the long range boundary dependency. Among the ice models the square lattice case has already revealed a striking phenomenon, the existence of an Arctic Circle delineating the random and ordered subdomains ([E]). Here we split the analysis for the other Archimedean lattices in two subsections since a surprising demarcation takes place. The material here contains both rigorous results and computer simulations using the principles from above.

\vskip .3truein
\noindent {\subtitle 4.1. Triangular and Kagom\'e cases}
\vskip .2truein

\noindent By the {\bf boundary height} of a configuration we mean the restriction of height to the boundary of the dual cover $D$ (see Section 1.): computing it we follow the boundary loop thereby only crossing boundary arrows.

Suppose that the boundary height is on each edge of the hexagon of constant tilt $0$ or $\pm 1.$ Let the {\bf signature} of the boundary be the six-vector that we get by recording the tilts starting from the base point and circumambulating the boundary of $D$ counterclockwise.

For both triangular and Kagom\'e lattice the signature $(+1,+1,0,-1,-1,0)$ and its cyclic permutations correspond to perfectly ordered configurations. They are frozen i.e. contain no directed 1-cycles.

The signature $(0,0,0,0,0,0)$ is the maximally disordered case and the configurations are temperate. Now the boundary arrows form unidirectional paths which are at least the length of the edge. The case where the entire boundary is a unidirectional cycle has this signature and clearly has the largest such cycle in any hexagon.

\vskip .2truein
\noindent Let the {\bf entropy of a boundary condition} for a $N$-hexagon denote the exponential size of the set of its legal fill-inns: $h_N={\log\left({\rm \#\ of\ configurations}\right)\over{\rm \#\ of\ arrows}}\ ,$ where $\#$ denotes the number on the entire domain. The {\bf entropy of a configuration} refers to the same number (any sample defines the boundary for all).

Suppose that the lattice spacing (the minimum of the distance between two neighboring lattice points) is set to be $1/N$. Then the configuration is defined in a discrete subset of a unit hexagon. If a sequence of scaled boundary heights $\{{1\over N}f_N\}$ converges as $N\rightarrow\infty$ to a limiting function $f$ we call the latter the boundary height of the scaling limit. Then $h=\lim_N h_N$ is the asymptotic exponential size of the set of legal configurations for this type of boundary condition. 

The entropy of the frozen case is obviously zero. Let the entropy of the maximally disordered case/free infinite model be denoted by $\overline h$

\proclaim Proposition 4.1.: For the triangular and Kagom\'e lattice the entropy can attain arbitrarily small positive values in the scaling limit. \par

\noindent {\bf Proof:}  Consider a seed configuration on the triangular lattice made of four subdomains with arrow orientations as indicated in Figure 5a.

The quadrant marked with $F$ is frozen; when the pca runs from the initial state illustrated the arrows in $F$ will remain unchanged. The opposite quadrant on the right is initially all directed 1-cycles (both even and odd). When the pca is run these will introduce directed 1-cycles to the top and bottom quadrants as well. 

Let $A_R$, $A_F^c$ and $A$ denote the number of 1-triangles in the right most quadrant, in the off-$F$ area and the total number respectively (essentially the numbers of arrows in these sets). By only flipping say even 1-triangles it is easy to establish the positive lower bound ${A_R\over 2A}\log 2$ for the entropy. Similarly the number ${A_F^c\over A}\overline h$ bounds the entropy from above (free action on $A_F^c$).

But the areas above are determined by the choice of cross point $x$, which in turn is determined by the boundary condition. If $x$ is the rightmost point on the boundary we have a zero entropy (frozen) case whereas if $x$ is the leftmost point on the boundary we are in the maximally disordered case. The boundary height is here of piecewise constant tilt hence the limiting boundary height trivially exists. Therefore we can reach arbitrarily small positive entropy value simply by the choice of $x$ (${\overline h}$ is finite). The construction for the Kagom\'e (and square) lattice is essentially the same and we omit it. \hfill\QED

\vskip .3truein
\centerline{\hbox{\psfig{figure=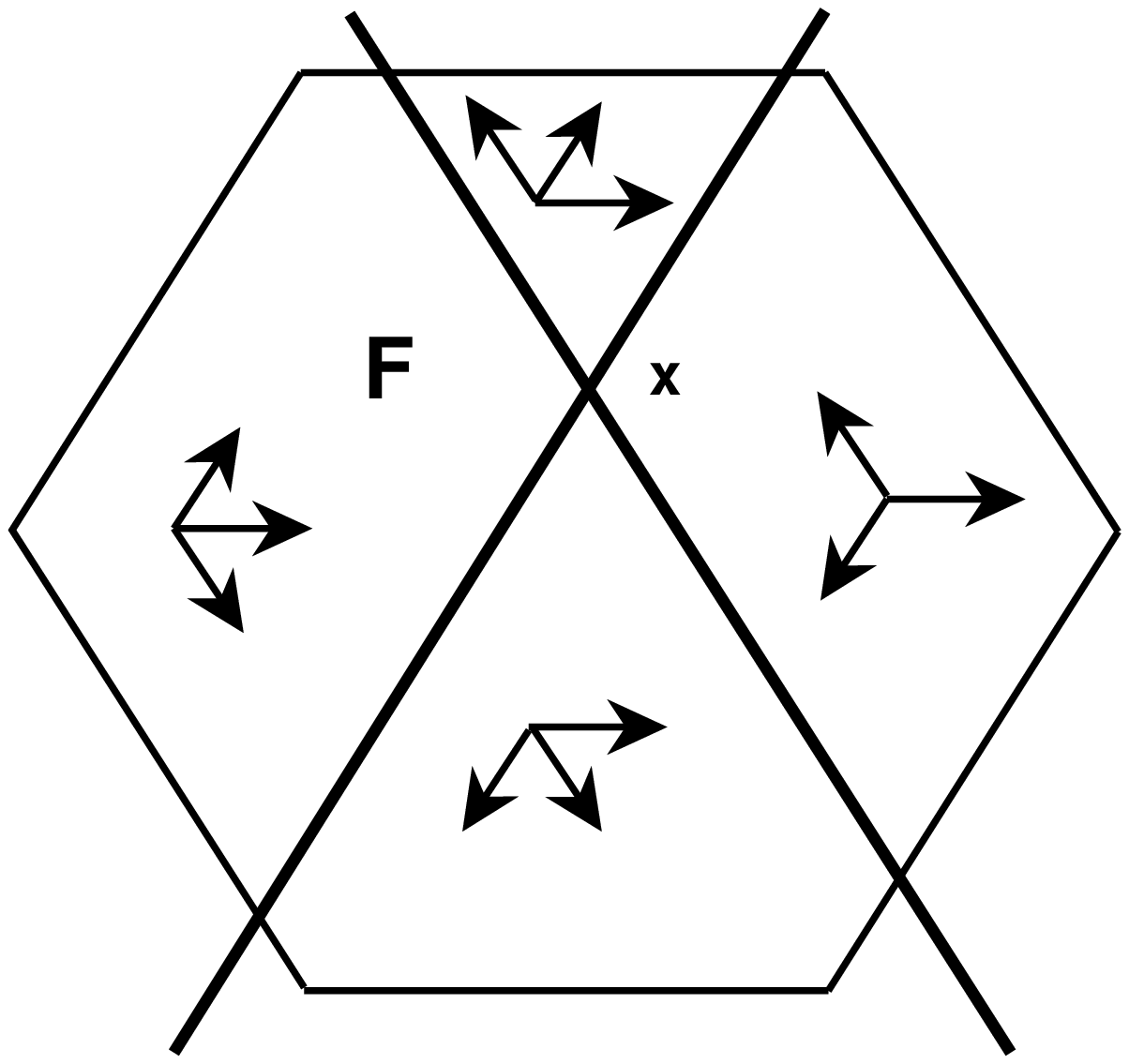,height=1.3in}
\hskip .7in \psfig{figure=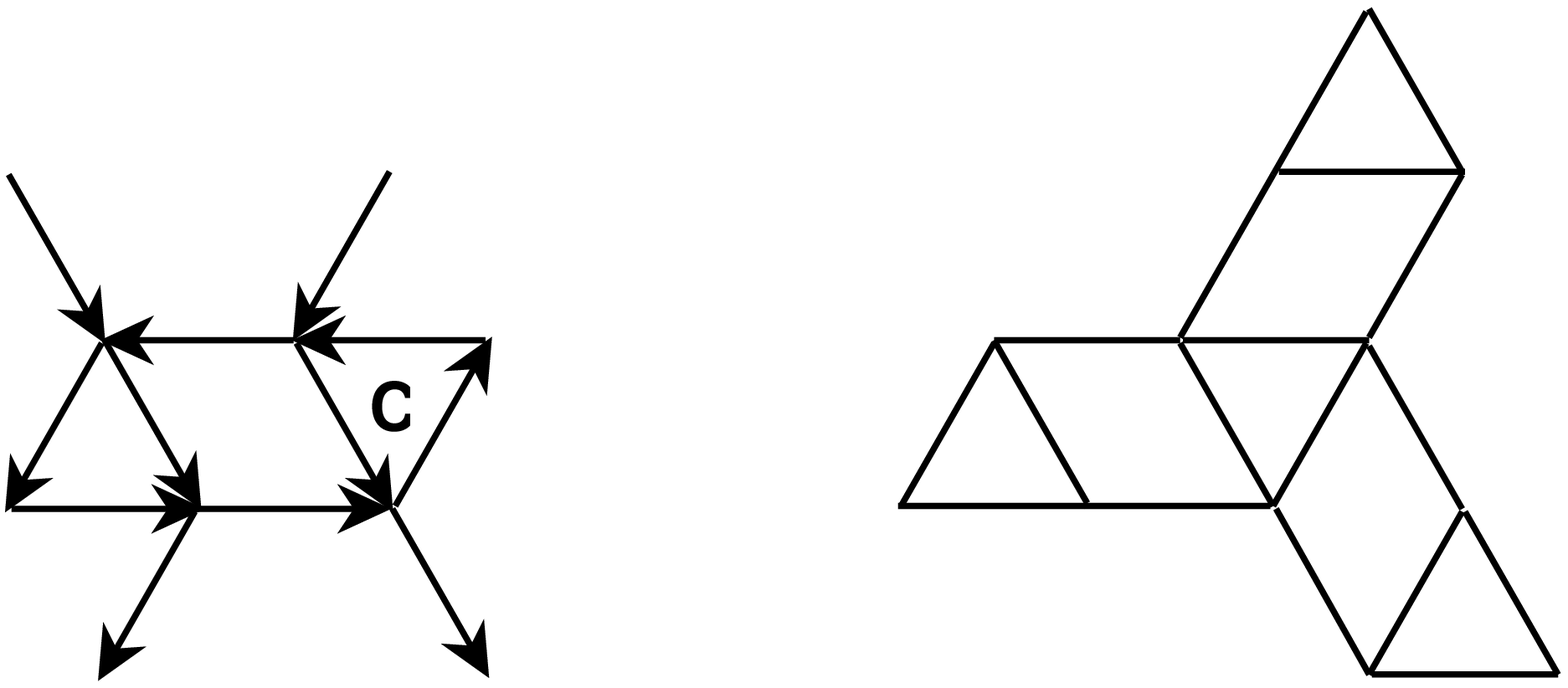,height=1.in} }  }
\nobreak
\vskip .2truein
\noindent {\bf Figure 5a.} Low entropy construction. {\bf b, c.} Forced cycle and the cycle template.

\vskip .2truein
\noindent A more general statement it likely to be true. However proving it would require some detailed information on the measure of the maximal entropy which we do not currently have.

\proclaim Claim 4.2.: For the triangular and Kagom\'e lattice the entropy can attain any value between 0 and $\overline h$ in the scaling limit. \par

\vskip .2truein
\noindent We now present some less trivial boundary conditions that lead to a striking illustration of the long range order in the ice on triangular and Kagom\'e set-up.

The signature choice $(+1,-1,+1,-1,+1,-1)$ is recorded in Figure 6a., top. This height choice leads to the coexistence of frozen and temperate domains. Figure 6a. middle illustrates the even 1-cycle flip distribution on triangular $102$-hexagon. Here we have recorded every even 1-cycle reversal during the iterates $13-19\times 10^3.$ The rendering is that of their cumulative totals during this period when the system was already close to equilibrium. Darker cells indicate higher flip activity. The light gray background is just to make the hexagon visible. There is a clear demarcation between corner areas where there is no activity and interior where the flip distribution is fairly uniform.

Figure 6b., top, shows another boundary condition on the same domain. Each edge is now split in the middle into a segments of extremal tilt $\pm 1$ in an alternating fashion. This choice freezes a lozenge shaped area at each corner (one of which is indicated). Figure 6b., middle, illustrates the cumulative even 1-cycle reversal distribution at equilibrium between iterates $6-12\times 10^3.$ In both this and the previous simulation the odd 1-cycles did show an indistinguishable distribution.

In the Kagom\'e case, too, these are perhaps the simplest boundary conditions yielding non-trivial interiors with symmetries. The bottom row of Figure 6. shows the cumulative cycle counts from iterates $12-24\times 10^3$ and $6-12\times 10^3$ for the two boundary choices (system again essentially at equilibrium). The coarser appearance of the images is due to the fact that the 1-cycle density is now half of that in the triangular case.

The rendering shows both even 1-triangle and 1-hexagon flips. The darker entries at the center indicate the array of triangles. At the center their flip frequency is about five times that for 1-hexagons. Note that if the arrows were laid down independently and with probability 1/2 to each orientation on a given edge, the ratio of the flip probabilities of an even 1-triangle to a 1-hexagon would be eight. Our pca here checked the 1-hexagons twice as often as even 1-triangles (with update cycle $\left\{F_e,F_h,F_o,F_h\right\}$). Hence if this distribution would be preserved it would generate flip probability ratio exactly four. So we can conclude that the maximally disordered vertex-configurations, the statistics of which we expect to see at the center, are some distance from uniform Bernoulli.

\vskip .2truein
\noindent While the fine structure of the interior in Figure 6. depends on the algorithm, the actual arrow distribution at the equilibrium does not. And most importantly the key result, the sharp demarcation of frozen and tempered subdomains akin to the Arctic Circle Theorem is plain in Figure 6. 

Due to the corner lozenges (in Figure 6b., top) one can actually compute upper bounds for the entropies on both lattice using the free models much the same way as was done in the square lattice case (for instance in Figure 6b the upper bounds are at most half of those for the free models).
 
%\vfill
\eject

\vskip .3truein
\noindent {\subtitle 4.2.$\ \ $ 3.4.6.4. lattice}
\vskip .2truein

\noindent The ice on the remaining Archimedean lattice, 3.4.6.4., turns out to be of quite different character. We don't know the ultimate reason for this - the combinatorics of the lattice just implies some odd properties.

One of them already lurks in Figure 2d. The figures attempted to illustrate the maximal boundary tilt in their lower halves. This succeeds for the two other lattices -  the heights are indeed a monotone increasing sequences downwards. But for 3.4.6.4. construction like that turns out to be impossible, revealed by the height value $2$ at the bottom.

Indeed there are no maximal tilt $\pm 1$ boundary conditions, hence there are no frozen configurations for ice on 3.4.6.4. The absolute value of the tilt is uniformly (in the size of the domain hexagon) bounded away from $1$. We conclude the presentation by formulating a \lq\lq no-go\rq\rq\ Proposition quantifying this and the entropy implications.

The lattice directions of the 3.4.6.4. lattice are the ones it inherits from the underlying triangular lattice ($0, \pm \pi/3$). Our hexagonal domain has its edges oriented along the lattice directions. Along these edges the 3.4.6.4. lattice viewed as a subset of the triangular lattice has period eight. An arrow block consisting of $n$ consecutive boundary arrows is called an $n$-{\bf block.}

\proclaim Proposition 4.3.: Consider a 3.4.6.4. configuration on a $N$-hexagon and an n-block along any of its straight edges. If the boundary arrows are of period 8 in the $n$-block the height over the block satisfies $|\Delta h|\le (3n+7)/4.$ For an arbitrary n-block of arrows, $n\ge 15$, the bound is $|\Delta h|\le (13n+28)/15.$ Hence if the boundary height exists in the scaling limit and has tilt, the absolute value of the latter cannot exceed 13/15. \hfill\break
In any 3.4.6.4. ice configuration in the set of 1-triangles and 1-lozenges at least $1/7$ of them are unidirectional. If the scaling limit entropy for a given boundary exist, it is bounded from below by ${1\over 24}\log2.$ 
\par

\noindent {\bf Proof:} The first statement follows from the observation that an all-in or all-out 8-block between two neighboring 1-hexagons immediately contradicts the ice rule in one of the boundary vertices. Therefore over such block the absolute height difference is at most 6. If a piece of the edge is made periodically of an 8-block, then in particular the 8-block between two 1-hexagons is an period block. By filling up a $n$-block with a maximal number of period blocks of length 8 we immediately obtain the first bound.

For the second bound we note that since a 15-block necessarily contains arrows from two 1-hexagons there must be at least one arrow of each orientation in such a block. Hence the height difference cannot exceed 13 in absolute value over the block. The filling argument over the $n$-block gives the stated bound which in turn implies the scaling limit bound once the tilt exists.

For the latter half of the statement pick a legal configuration and a 1-triangle in it. Suppose that it isn't unidirectional. Up to rotation and reflection it will look like the triangle on the left of Figure 5b. Then either the lozenge on its right is unidirectional or if it isn't, the triangle next to it on the right must be. Hence in any Y-shaped arrangement of 1-triangles and 1-lozenges (as in Figure 5c.) there is at least one 1-cycle. As there are at most 24 arrows in this Y-plaquette determined by the construction, the lower bound follows. \hfill\QED

\vskip .2truein

\noindent Recall that besides having completely frozen configurations, by Proposition 4.1. the ice on triangular and Kagom\'e lattices can have arbitrarily low entropy configurations (and so can square ice by similar construction). By the Proposition above the situation on 3.4.6.4. differs on both counts and implies that configurations are of quite different character.

The height and entropy bounds of the Proposition are likely not to be tight. Indeed it seems rather difficult to design \lq\lq stiff\rq\rq\ configurations of any kind. The lowest entropy boundary configurations that we have been able to construct have entropy ${1\over 6}\log2$. The seed for such configuration was illustrated in Figure 4d. The fattened hexagon can be extended periodically to an arbitrarily large, unique configuration. In this configuration all the horizontal right leaning parallelograms like the one with bold arrows in the Figure 4d. can generate under the pca action exactly three other local arrow arrangements. Accounting the density of these parallelograms in the configuration immediately gives the entropy value.

The highest entropy boundary condition that we know of (its seed configuration having all 1-cycles directed) has the entropy at least ${1\over 4}\log2$ (which thereby is a lower bound for the entropy of the free model on 3.4.6.4.).

\vskip .2truein
\noindent In view of the results it should come as no surprise that the configurations of 3.4.6.4. ice look disordered and rather homogeneous for any boundary condition. Experimenting with hexagons of size around $N=100$ we found faint traces of boundary dependency in the statistics of the interior (e.g. slight variation in the 1-cycle flip densities). But since frozen states do not exist for 3.4.6.4. there is no possibility of such clear demarcation result as the Arctic Circle/Flower exhibited by the ice on the other Archimedean lattices.

\vskip .4truein
\noindent {\subtitle References}
\vskip .2truein

\item{[B]} Baxter, R.J.: {\sl Exactly solvable models in statistical mechanics},
Academic Press, 1982.
%\item{[vB]} van Beijeren, H.: Exactly solvable model for the roughening transformation of a crystal surface, {\sl Phys. Rev. Lett.}, {\bf 38}, No. 18 (1977), 993-996.
\item{[CL]} Conway, J.H., Lagarias, J.C.: Tilings with polyominoes and combinatorial group theory, {\sl J. Combin. Theory}, Ser. A, {\bf 53} (1990), 183-208. MR1041445 (91a:05030)
\item{[E]} Eloranta, K.: Diamond Ice, {\sl J. Stat. Phys.} {\bf 96} 5/6 (1999), 1091-1109. MR1722988 (2000f:82014)
\item{[GS]} Gr\"unbaum, B., Shephard, G. C.: {\sl Tilings and patterns}, Freeman, 1987.
\item{[JPS]} Jockush, W., Propp, J., Shor, P.: Random domino tilings and the arctic circle Theorem, {\tt arXiv:math.CO/9801068}. 
\item{[K]} Kasteleyn, P.: The statistics of the dimer on a lattice, I. The number of dimer arrangements on a quadratic lattice, {\sl Physica} {\bf 27} (1961), 1209-25.
\item{[KZ-J]} Korepin, V., Zinn-Justin, P.: Thermodynamic limit of the Six-Vertex Model with Domain Wall Boundary Conditions, {\sl J. Phys.} {\bf A 33} No. 40 (2000), 7053, {\tt arXiv:cond-mat/0004250v4}.
\item{[L]} Lieb, E.H.: Residual entropy of square ice, {\sl Phys. Rev.} {\bf 162} (1967), 162-172.
\item{[P]} Propp, J.: Lattice structure for orientation of graphs, {\tt arXiv:math.CO/0209005}. 
\item{[T]} Thurston, W.P.: Conway's tiling groups, {\sl Am. Math. Monthly} (Oct. 1990), 757-773. MR1072815 (91k:52028)
\item{[W]} Walters, P.: {\sl An Introduction to Ergodic Theory}, Springer, 1982.

\vfill
\eject

\voffset 1truein
\centerline{\hskip .25truein
\hbox{\psfig{figure=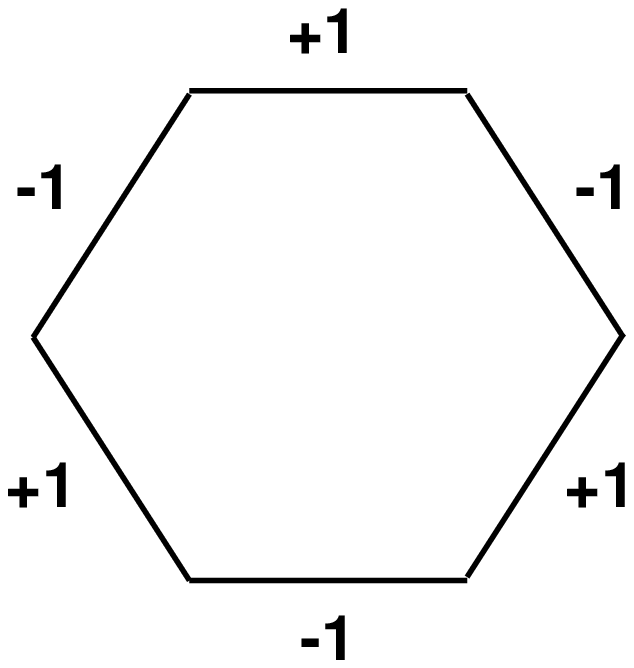,height=2.1in} 
\hskip .5in\psfig{figure=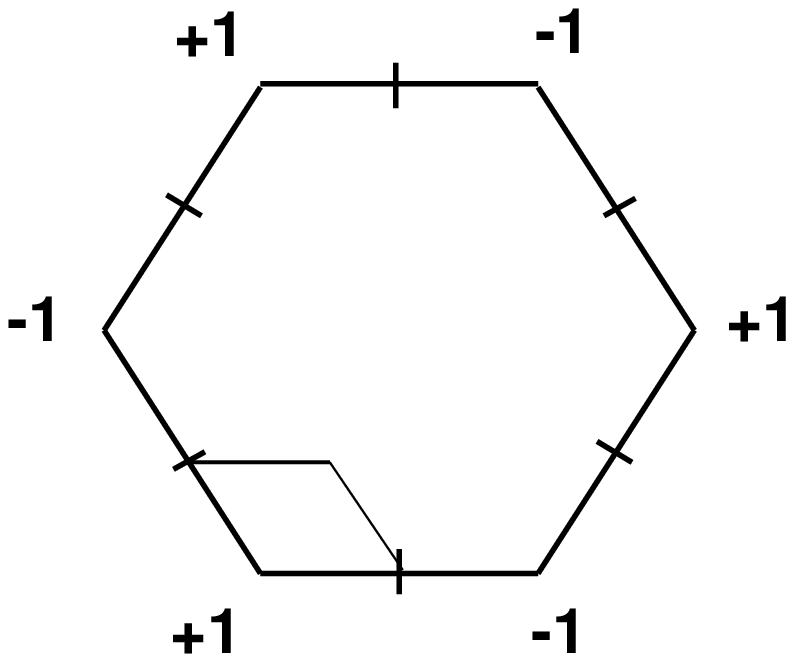,height=2.1in} } }
\vskip .1truein
\centerline{\hbox{\psfig{figure=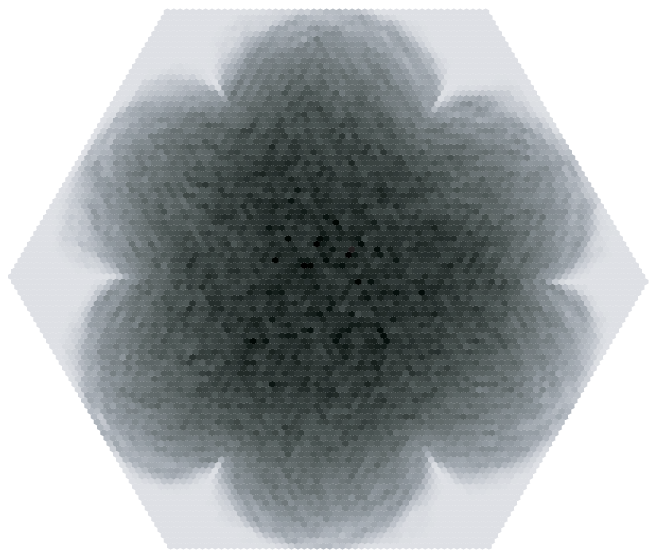,height=1.55in} 
\psfig{figure=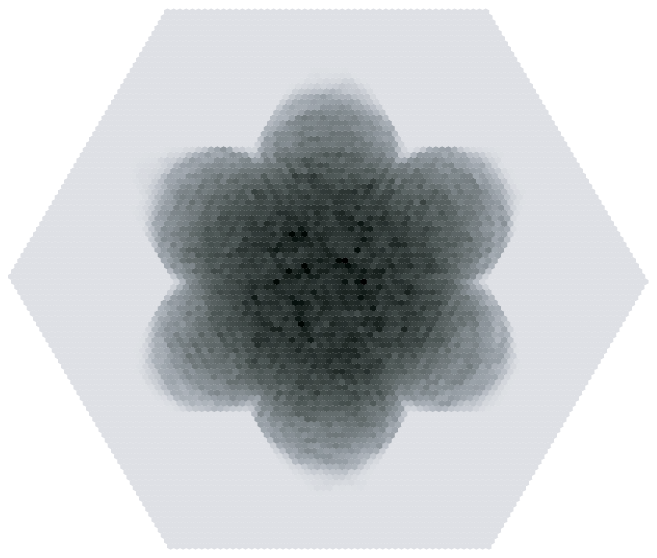,height=1.55in} } }
\vskip .3truein
\centerline{\hbox{\psfig{figure=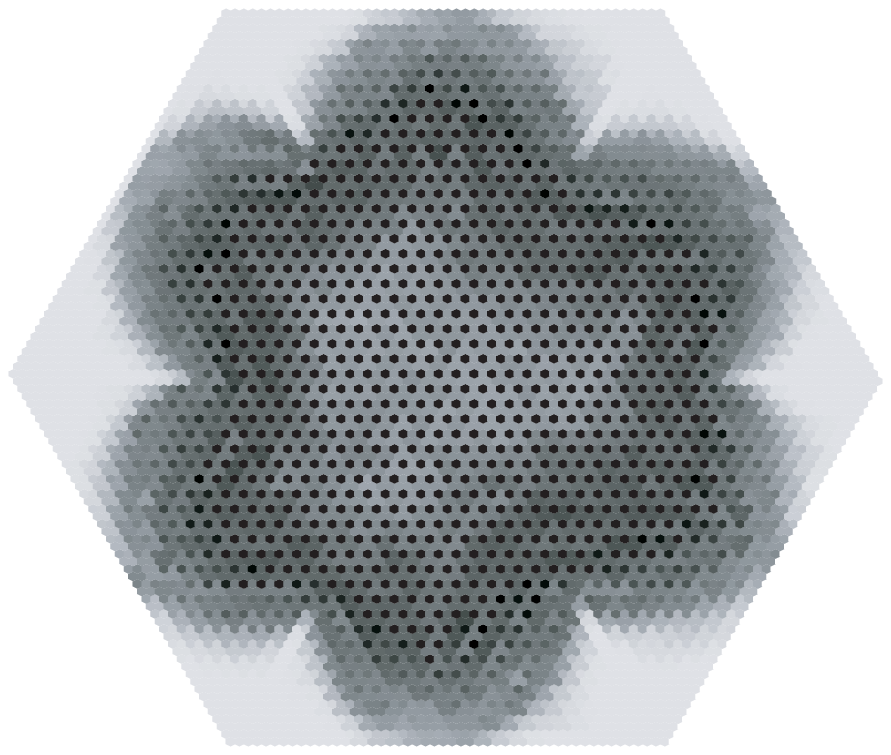,height=1.55in} 
\psfig{figure=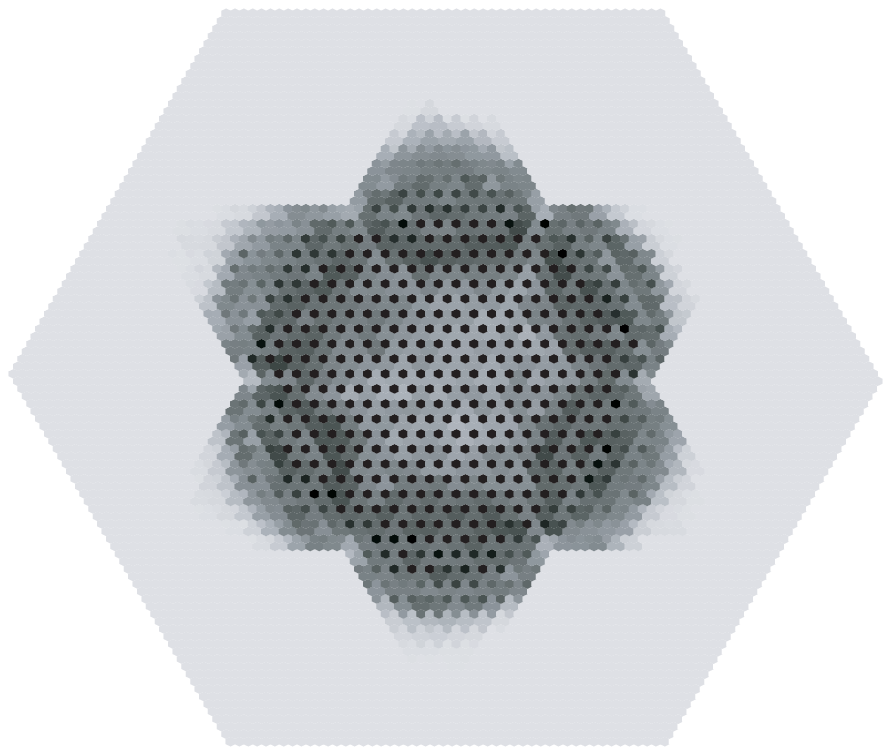,height=1.55in} } }

%\centerline{\hskip .25truein
%\hbox{\psfig{figure=zigzag1.eps,height=2.1in} 
%\hskip .5in\psfig{figure=david1.eps,height=2.1in} } }
%\vskip .1truein
%\centerline{\hbox{\psfig{figure=19k2bm.eps,height=1.55in} 
%\psfig{figure=12k5bm.eps,height=1.55in} } }
%\vskip .3truein
%\centerline{\hbox{\psfig{figure=6k2bm.eps,height=1.55in} 
%\psfig{figure=3k1bm.eps,height=1.55in} } }
\vskip .5truein
\noindent {\bf Figure 6a, b (left and right columns).} Top: domain with boundary signatures, middle: triangular, bottom: Kagom\'e lattice configurations.

\vfill
\eject
\end